\documentclass[fleqn,usenatbib]{mnras}

\usepackage{newtxtext,newtxmath}

\usepackage[T1]{fontenc}

\DeclareRobustCommand{\VAN}[3]{#2}
\let\VANthebibliography\thebibliography
\def\thebibliography{\DeclareRobustCommand{\VAN}[3]{##3}\VANthebibliography}

\usepackage{mathtools}

\usepackage{graphicx}	
\usepackage{amsmath}	
\usepackage{hyperref}
\usepackage{xcolor}

\title[Persistent magnetic cycles during grand minima]{Evidence of persistence of weak magnetic cycles driven by meridional plasma flows during solar grand minima phases}

\author[Saha et al.]{
Chitradeep Saha,$^{1}$
Sanghita Chandra,$^{1,2}$
and Dibyendu Nandy$^{1,2}$\thanks{E-mail: dnandi@iiserkol.ac.in}
\\
$^{1}$Center of Excellence in Space Sciences India, Indian Institute of Science Education and Research Kolkata,
Mohanpur 741246, West Bengal, India\\
$^{2}$Department of Physical Sciences, Indian Institute of Science Education and Research Kolkata,
Mohanpur 741246, West Bengal, India\\
}

\date{Accepted XXX. Received YYY; in original form ZZZ}

\pubyear{2022}

\begin{document}
\label{firstpage}
\pagerange{\pageref{firstpage}--\pageref{lastpage}}
\maketitle

\begin{abstract}
 
Long-term sunspot observations and solar activity reconstructions reveal that the Sun occasionally slips into quiescent phases known as solar grand minima, the dynamics during which is not well understood. We use a flux transport dynamo model with stochastic fluctuations in the mean-field and Babcock-Leighton poloidal field source terms to simulate solar cycle variability. Our long-term simulations detect a gradual decay of the polar field during solar grand minima episodes. Although regular active region emergence stops, compromising the Babcock-Leighton mechanism, weak magnetic activity continues during minima phases sustained by a mean-field $\alpha$-effect; surprisingly, periodic polar field amplitude modulation persist during these phases. A spectral analysis of the simulated polar flux time series shows that the 11-year cycle becomes less prominent while high frequency periods and periods around 22 years manifest during grand minima episodes. Analysis of long-term solar open flux observations appears to be consistent with this finding. Through numerical experimentation we demonstrate that the persistence of periodic amplitude modulation in the polar field and the dominant frequencies during grand minima episodes are governed by the speed of the meridional plasma flow -- which appears to act as a clock.  
 
\end{abstract}




\section{Introduction}

Direct observations of solar activity in the past complemented by multi-millennium reconstructions of its different proxies \citep{usoskin2007grand, usoskin2014evidence, usoskin2017history} reveals that our host star -- the Sun -- can sometime slip into a quiescent phase known as the solar grand minimum. The last such grand minimum in solar activity was recorded during 1645-1715 and is known as the Maunder minimum. These episodes are characterised by reduction or absence of sunspots on the solar surface. These strongly magnetised dark spots \citep{hale1908zeeman} serve as our main proxy to understand solar activity. The Sun is our primary source of energy and its activity modulates our space environment, space-based technologies and planetary atmospheres over short-to-long timescales \citep{SCHRIJVER20152745, Nandy2021}. Grand minima are extreme activity phases accompanied by significant reduction in solar radiative, particulate and magnetic output with consequences for the state of the heliosphere. It is therefore important to understand the dynamics of solar activity during these phases.

The Sun is believed to have a dynamo mechanism operating in its interior. This sustains the generation and recycling of magnetic fields. The flux ropes generated within the solar convection zone (SCZ) become buoyant and reach the surface to generate bipolar sunspot regions \citep{parker1955hydromagnetic}. Sunspots on the photosphere serve as an indicator of magnetic field generation deep within \citep{nandy2002explaining, 2002Ap&SS.282..209N}. However, when the photosphere is devoid of sunspot activity, it becomes difficult to constrain solar magnetic field dynamics in its interior. Herein solar dynamo modeling plays a crucial role in probing solar internal magnetic field dynamics. From the theoretical perspective it is important to understand what drives the Sun into such quiescent phases, what induces recovery, and what is the nature of solar activity during grand minima phases? This outlines the motivation for our study with a specific focus on the last question.

Through flux transport dynamo modelling incorporating stochastically fluctuating mean-field \citep{parker1955hydromagnetic, 1961ApJ...133..572B, 1969ApJ...156....1L} poloidal field sources we simulate solar grand minima like episodes. We then look at trends in the polar flux at the surface and toroidal flux buried deep within to understand the underlying dynamics during these phases. We find persistent weak magnetic cyclic activity in the flux and identify certain trends. The first observational evidence of cyclic solar magnetic activity during Maunder minimum was put forward by  \cite{Beer1998} through a measurement of $^{10}$Be concentration in the Arctic ice core. Our analyses of the simulated polar flux time series identify signatures of a prominent component around $\thicksim$5-year over and beyond a persisting 11 year periodicity during grand minima. There have been hints of such high frequency cycles during regular periods as well in solar activity observations \citep{Kane2007, Zi_ba_2006}. We conclusively establish that the frequency of these weak, persistent cycles during grand minima episodes is governed by the speed of the meridional plasma flow in the solar interior.   

\section{Model setup} \label{sec: methods}

\begin{figure}
    \centering
    \begin{minipage}{0.48\textwidth}
        \centering
        \includegraphics[width=0.8\textwidth]{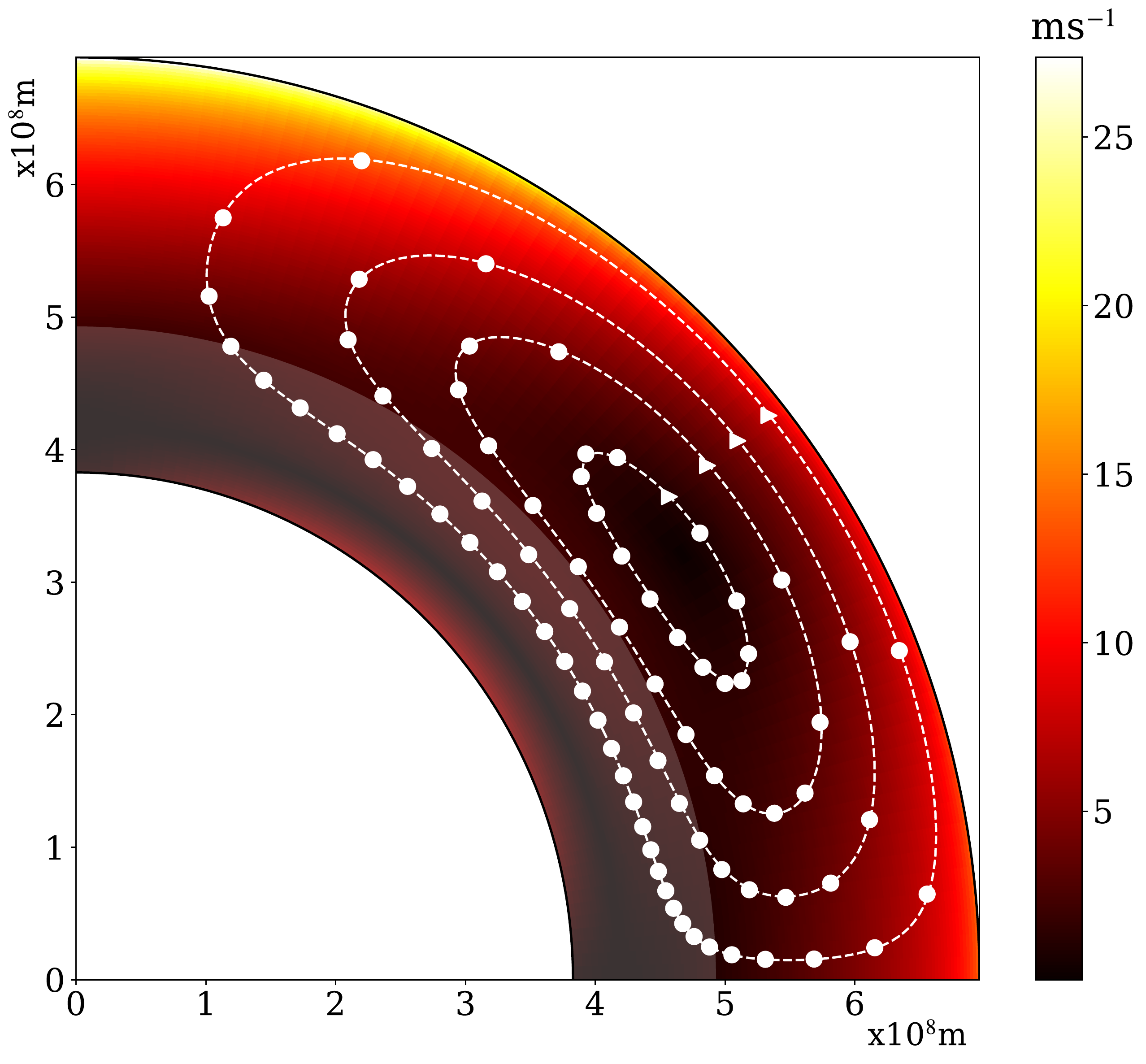} 
    \end{minipage}\hfill
    \caption{Meridional circulation profile used in our model. Dots on the streamlines show yearly positions for imaginary particles advected with peak flow speed of 29 ms$^{-1}$, starting from the triangles at $45^{\circ}$ north  and moving anticlockwise.}
    \label{fig:meridional}
\end{figure}

We utilise an axisymmetric dynamo model -- \texttt{SURYA}, working in the kinematic regime \citep{Nandy_2001, chatterjee2004full}. The global magnetic field is expressed as a combination of the following components in spherical polar coordinates:
\begin{equation}
    \mathbf{B} = B(r, \theta)\mathbf{e_{\phi}}+\mathbf{B_p}
\end{equation}
where $\mathbf{B_p}$ = $\mathbf{\nabla} \times[A(r, \theta) \mathbf{e_{\phi}}$] gives the poloidal component and $B(r, \theta)\mathbf{e_{\phi}}$ denotes the toroidal component. The dynamo equations obtained after solving the magnetic induction equation for \textbf{B} govern the evolution of these components:
\begin{equation}
    \frac{\partial A}{\partial t} + \frac{1}{s}\left[ (\mathbf{v_p} \cdot \nabla) (sA) \right] = {\eta}_p\left( \nabla^2 - \frac{1}{s^2}  \right)A + \alpha B
\end{equation}

\begin{equation}
\begin{split}
    \frac{\partial B}{\partial t}  + s\left[ \mathbf{v_p} \cdot \nabla\left(\frac{B}{s} \right) \right] + & (\nabla \cdot \mathbf{v_p})B  \\ = {\eta}_t\left( \nabla^2 - \frac{1}{s^2}  \right)B 
    &  +  s\left(B_p\cdot \nabla \Omega\right)   + \frac{1}{s}\frac{\partial (sB)}{\partial r}\frac{\partial {\eta}_t}{\partial r}
\end{split}
\end{equation}

\noindent where s = r sin $\theta$.  Meridional flow is denoted by $\mathbf{v_p}$, whereas $\Omega$ represents the rotation in the solar convection zone (SCZ). The symbols $\eta_p$, $\eta_t$ and $\alpha$  denote the poloidal and toroidal diffusivities and source term for the poloidal field respectively.  Our model has a single cell meridional circulation profile in each hemisphere \citep{nandy2002explaining, chatterjee2004full}. The flow profile is designed to capture a faster poleward flow and slower equatorward counterflow (see Fig. \ref{fig:meridional}). The rate at which fluid particles are advected along the flow streamlines varies with the peak flow speed.

We use the mean field $\alpha$-source $\left(\alpha_{MF}\right)$, operating in the bulk of the SCZ, and the Babcock-Leighton $\alpha$-source $\left(\alpha_{BL}\right)$, operating near the surface. Stochastic fluctuations are inserted independently in both the hemispheres. The idea behind introducing fluctuation in $\alpha_{BL}$ is to mimic the variation in efficiency of the surface BL mechanism towards magnetic bipoles of varied tilt angles \citep{passos2014solar}. Likewise, $\alpha_{MF}$ fluctuates to capture the effect of turbulent helical convection deep within the SCZ. The turbulent buffeting of the buoyant magnetic flux tubes rising through the SCZ adds a random component to the tilt angle, contributing to a dispersion in the tilt angle  distribution. On the other hand the weak mean-field $\alpha$ effect $\alpha_{MF}$ operates in the bulk of SCZ on weak flux tubes that are below a certain threshold and which do not contribute to the formation of sunspots. Thus $\alpha_{MF}$ effect, unlike $\alpha_{BL}$, does not cease its operation even during solar grand minima phases. $\alpha_{MF}$ working in conjunction with the surface $\alpha_{BL}$ facilitates the Sun to recover from its quiescent phases \citep{passos2014solar, Hazra_2014, 10.1093/mnrasl/slab035}.

\begin{figure*}
    \centering
    \includegraphics[width=0.95\textwidth, height = 8cm]{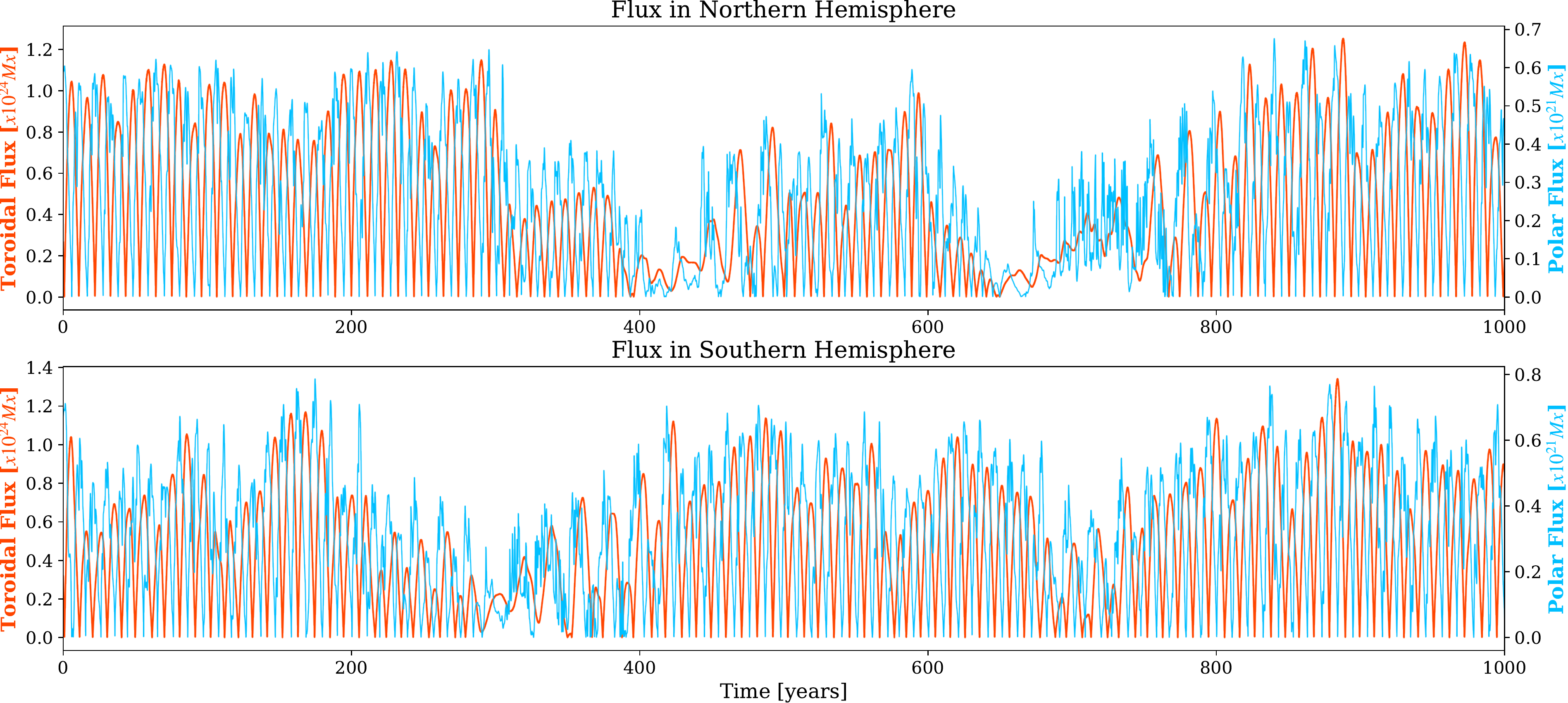}
    \caption{Simulated time series for the hemispheric polar flux [blue] emerging out of the high latitude solar surface regions and the toroidal flux [red] penetrating the meridional plane at the vicinity of the tachocline. Top (bottom) panel denotes the fluxes in the northern (southern) hemisphere. }
    \label{fig:fluxseries}
\end{figure*}

\begin{figure*}
    \centering
    \includegraphics[width=0.95\textwidth, height = 8cm]{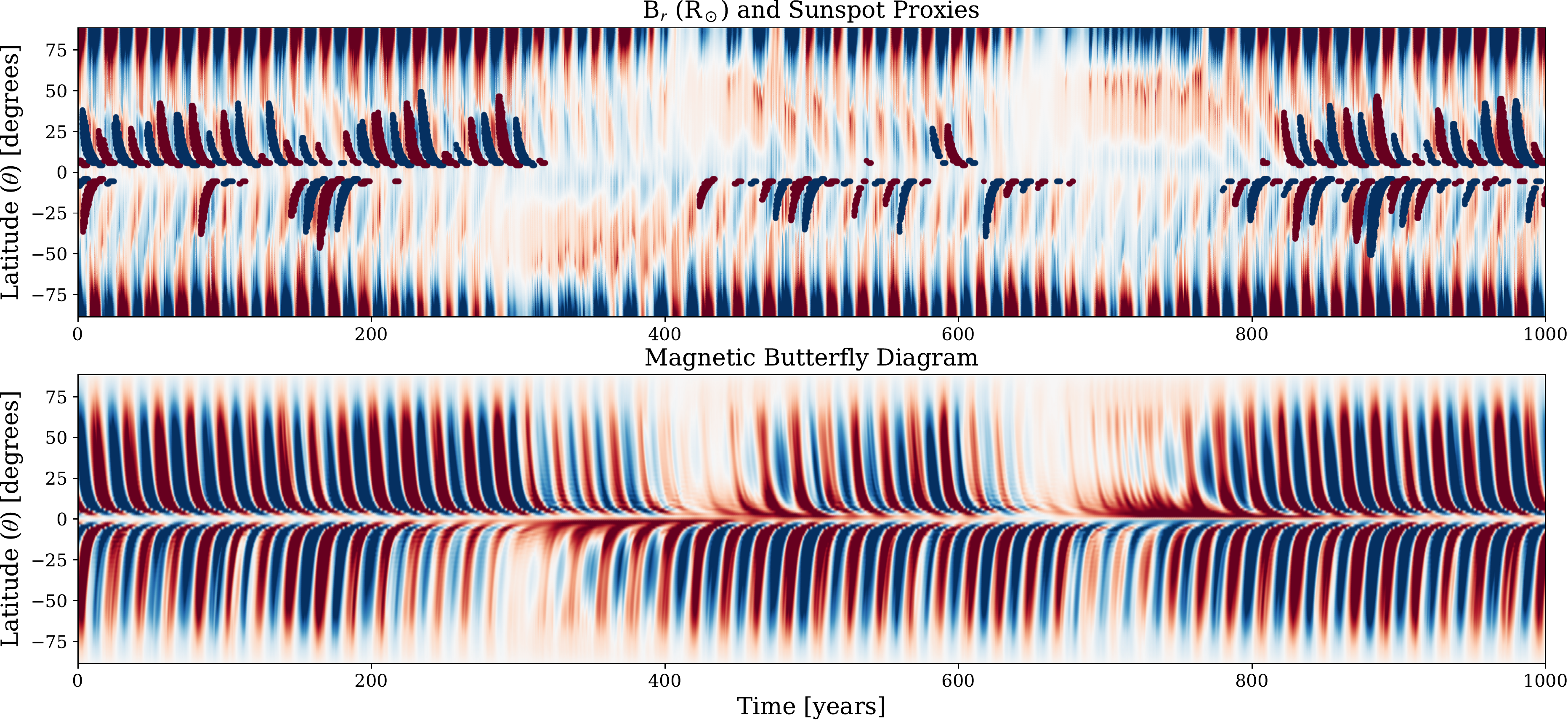}
    \caption{Top panel: simulated time series of the  radial magnetic field (saturated to 1500G) with the sunspot eruptions primarily confined to lower latitudes. Bottom panel: toroidal magnetic field (saturated to 150kG) at the base of the convection zone .}
    \label{fig:butterfly}
\end{figure*}

In our model, additive stochastic fluctuations are introduced to the mean values ($\alpha_{BL}^0$ and $\alpha_{MF}^0$) in poloidal source profiles independently in both the hemispheres as $\alpha_{BL}$ = $\alpha_{BL}^0$ + $\alpha_{BL}^0$ $\sigma_1$(t,$\tau_1$) and $\alpha_{MF}$ = $\alpha_{MF}^0$ + $\alpha_{MF}^0$ $\sigma_2$(t,$\tau_2$). The second term in both the expressions is the fluctuating term, where $\sigma_1$ is a random number uniformly varying in the range [-1.5,1.5] and $\sigma_2$ varies in [-1,1]. $\tau$ denotes the coherence timescale for the stochastic fluctuations. In the present study we use $\alpha_{BL}^0 = 27$ m/s (below 21 m/s solutions decay) and $\alpha_{MF}^0 = 0.4$ m/s, $\tau_1 =$ 6 months and $\tau_2 =$ 1 month. All other input profiles are adapted from \cite{passos2014solar}.

\begin{figure*}
 \centering{\includegraphics[width=0.8\textwidth,clip=]{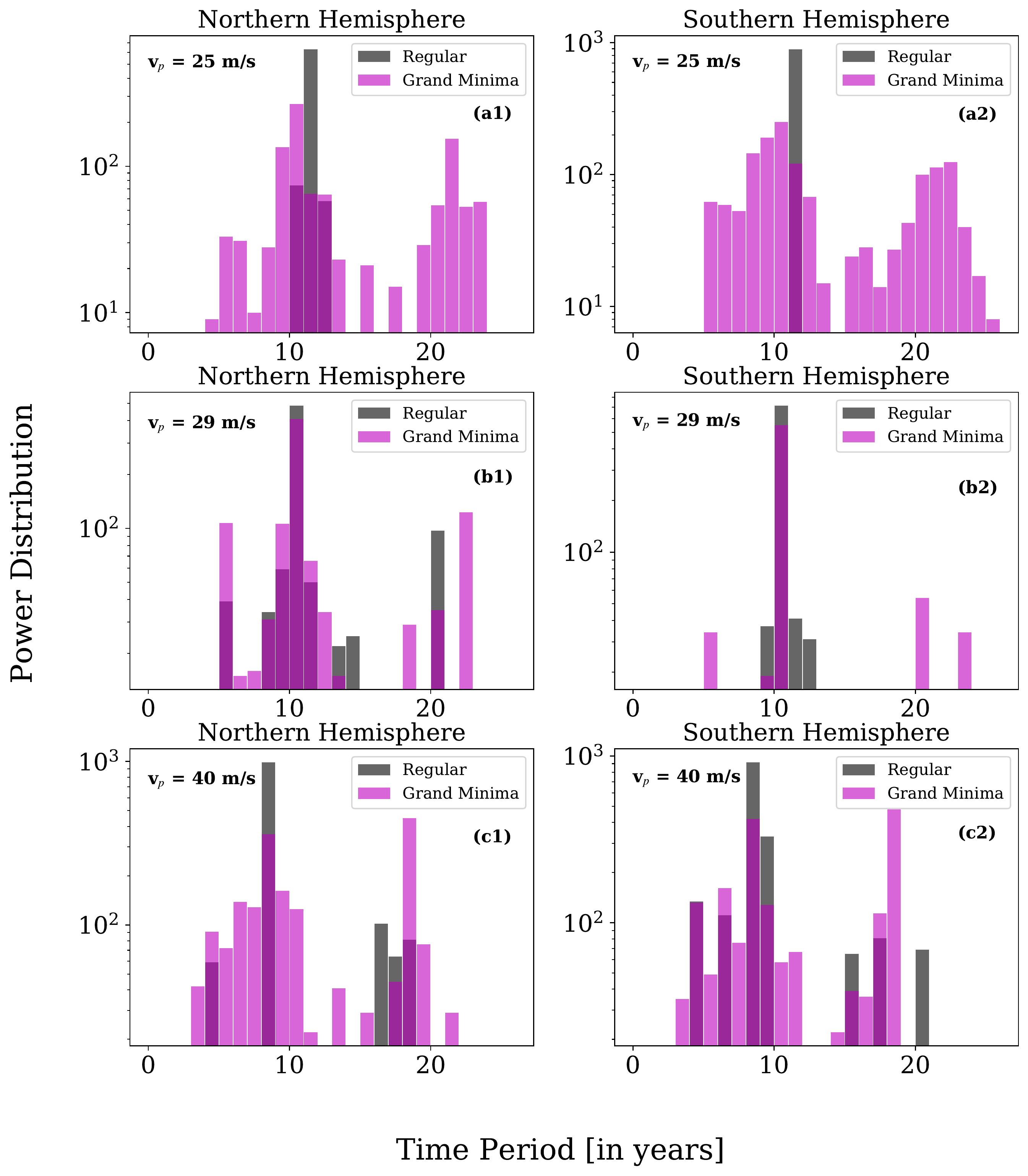}}
\caption{Cumulative power distribution of the dynamo simulated hemispheric polar flux time series during regular solar activity and grand minima epochs, for three different peak flow speeds of the meriodional circulation,  a1-2) 25 m/s [upper panel], b1-2) 29 m/s [middle panel], c1-2) 40 m/s [lower panel]. Spectral power appears to be redistributed across higher and lower frequencies during grand minima phases. The spectra as a whole shift towards the higher frequency i.e. lower time period regime as the peak flow speed increases and vice-versa.} \label{Fig: nutshell}
\end{figure*}

\section{Results}\label{sec: results}
The level of stochasticity in the poloidal sources plays an important role to impact the cycle amplitude. Furthermore, the coherence timescales  $\tau_1$ and $\tau_2$ are typically much smaller than the decadal solar cycle timescale. Therefore, multiple perturbations take place during 11-year solar cycles. Consecutive events with positive perturbation signify addition to the polar field source from active regions which follow the Joy's law and Hale's polarity law, whereas consecutive negative perturbations can potentially drive the system to a magnetically dormant state with no significant eruption events. For more complex analysis on the impact of poloidal source fluctuations at different phases of magnetic cycles see \cite{2018A&A...615A..38K}. If the dormancy persists for at least three consecutive solar cycles we consider it to be a grand minimum. It is to be noted that the stochastic forcing in our simulations results in hemispheric asymmetry which manifests through asynchronous occurrence of hemispheric grand minimum. Nevertheless our long term dynamo simulation spanning over ten millennia produced as many as 22 grand minima episodes with temporal overlapping in both the hemispheres -- about 2.4$\%$ of all the solar cycles -- with a distribution of duration ranging from 40 years to 180 years, a statistics that compares reasonably well with observations and reconstructions \citep{usoskin2007grand}.

The simulations reveal certain sequence of events at the onset and during the grand minima -- a decay in the polar flux and in turn the toroidal flux amplitudes, followed by occasional halts in the polar field reversal (see Fig. \ref{fig:fluxseries} \& Fig. \ref{fig:butterfly}). The buoyancy algorithm in our model facilitates sunspot eruptions only when the toroidal field is stronger than 80 kG. Therefore, the decaying toroidal field amplitude results in the disappearance of the sunspots (see Fig. \ref{fig:butterfly}, top panel). Polar flux continues to drop until it reaches a minimum value, after which the polar field reversal stops due to insufficient surface field strength. Interestingly, the weak toroidal field seated deep in the convection zone continues to flip its polarity during this period (see Fig. \ref{fig:butterfly}, bottom panel). This implies that the large scale plasma flows --  along with the turbulent diffusion -- are still operational within the SCZ and continues to drive a weak internal dynamo. These transport mechanisms dredge the toroidal field up to the SCZ, where the MF poloidal source acts to generate polar fields. This contributes to field build up and concentration at the polar caps due to the poleward meridional flow. It is to be noted that the dominant polarity field accumulated there is same as the polarity of the last significant sunspot cycle before entering into a grand minimum.

The toroidal and polar flux evolution serve as good proxies to probe the internal dynamics. In our simulated data, it is observed that with the decay of toroidal flux beneath a threshold, comes the stoppage of sunspot eruptions. The polar flux also decays in this phase. A particularly intriguing phenomenon is noticed in the behavior of polar flux; fluctuations with time periods shorter than the decadal timescale become apparent (see Fig. \ref{fig:fluxseries}). We perform spectral analysis on the hemispheric flux to find that these timescales correspond to approximately 5 years (Fig. \ref{Fig: nutshell}(b)). Previous studies have established that it takes about 5 years for toroidal field at the base of the tachocline to be dredged up to the surface \citep{passos2014solar}. Whether this correspondence is random or there exist causal connections is what we explore next. We also check for any significant systematic changes in the mean field $\alpha$ and note that there are no noticeable trends in its nature during grand minima episodes.

We perform spectral analysis using fast Fourier transform method and impose a power threshold in order to eliminate physically irrelevant Fourier overtones (of the fundamental 11 year cycle) along with other insignificant frequencies appearing due to the stochastic forcing in our dynamo model. Two separate cases are considered: case 1, when spectral analysis is performed on the regular cyclic phases of simulated data, and case 2, when the same is done on the grand minima phases. The distribution of spectral intensities over multiple phases in both cases 1 and 2, give information about the relative power of the different frequencies that are present. Our findings suggest that the power stored in the time period of about 11 years is fairly high in both the regular and grand minima epochs. However, when comparing the power stored in the 11 year period and shorter time periods of around 5 years, the relative power stored in the shorter time periods become more pronounced for case 2. Curiously, there is also an enhancement in spectral power around $\sim$22 year period, which we attribute to the fact that the last dominant polarity of the polar field before entry in to grand minima phases dominates with a jump of one cycle during the low activity mode.  

It is further seen that with increasing peak flow velocities in the meridional circulation, the power spectrum shifts to lower time periods (higher frequencies) and vice versa (see Fig. \ref{Fig: nutshell}). This establishes that meridional circulation is the primary determinant of the periodicities during grand minima episodes.

Various flux transport processes combine to govern the dynamics of the solar cycle. Meridional circulation, one of the slowest processes, regulates the solar cycle period. When eruptions stop, weak magnetic cycles still persist in the solar interior implicating that the magnetic processes do not halt completely, possibly due to the mean field alpha effect \citep{passos2014solar, Hazra_2014}. The 11 year periodicity primarily arises from sunspot eruptions. However, during grand minima epochs, eruptions cease in our simulations. A closer look at the flux evolution indicates that cycles weak in magnitude continue; but the 11 year period becomes less prominent relative to regular phases. The high frequency cycles associated with the meridional circulation advecting toroidal fields through the SCZ, where the poloidal field is concentrated, then becomes important. During the regular solar cycles, these higher frequencies (sustained by mean field alpha effect) are masked by the dominating 11 year cycle period. The $\thicksim$ 5 year cyclicity is apparent only in the polar flux which further supports our hypothesis of meridional circulation being responsible for it as it is the circulation which aids in polar field build up. The toroidal flux do not show such prominent high frequency cycles. This is supported by theoretical studies of meridional circulation profiles which indicate that the equatorward counterflow corresponds to a timescale of $\thicksim$ 12 years \citep{nandy2004meridional}.

\begin{figure}
    \centering
    \begin{minipage}{0.37\textwidth}
        \centering
        \includegraphics[width=\textwidth]{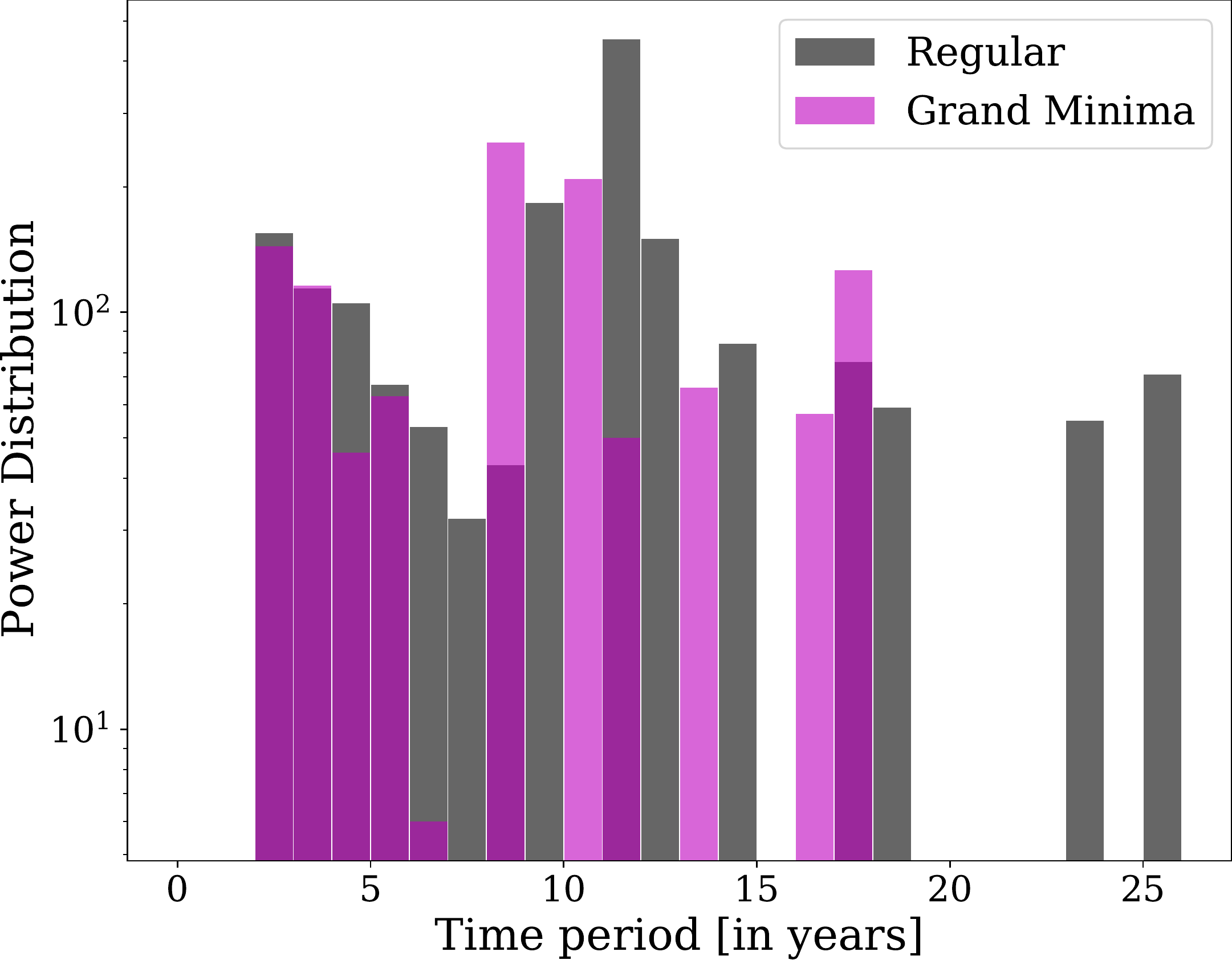} 
    \end{minipage}
    \caption{Distribution of relative power stored in different time periods in the annually resolved millennium scale, reconstructed open solar flux time series.}
    \label{fig:actual_recon_wiki}
\end{figure}

We further analyze millennium scale open solar flux time-series \citep{2021yCat..36490141U} which is reconstructed from $^{14} {C}$ data \citep{usoskin2014evidence} and which serves as a good proxy for the global solar polar flux. Although, given the lower time extent and lower resolution ($\sim 1$ year) of the reconstructed time-series, the peaks are not that significant, as far as trends are concerned, we find a similar reduction in the power in the 11 year period and redistribution across other periodicities (see Fig. \ref{fig:actual_recon_wiki}). An independent work simulating long-term cosmic ray modulation potential due to solar dynamo activity finds similar trends \citep{2022arXiv220812103D}

\section{Concluding Remarks}\label{sec: discussion}

Using dynamo simulations of solar activity we explore the dynamics of magnetic fields in the Sun's interior during simulated grand minima episodes. Based on our findings, we conclude that the meridional circulation in the Sun's interior and a mean field $\alpha$-effect in the solar convection zone can sustain weak, magnetic cycles in the large-scale polar field amplitude even during solar grand minimum.  This reveals that some facets of the underlying dynamics in the solar interior continue during these magnetically quiescent phases. Analysis of solar open flux reconstruction hints at the presence of similar periodic trends, lending independent support to our results.

In our simulations we find that the speed of the meridional circulation governs the periodicities manifest in the solar activity during grand minima phases. Therefore, it appears that the meridional plasma flow threading the solar convection zone continues to function like a clock during solar grand minimum, and the signature of this process is manifest even in the absence of regular sunspot cycles.

\section*{Acknowledgements}

The authors thank Soumyaranjan Dash, Shaonwita Pal and the anonymous referee for useful comments. C.S. acknowledges financial support from CSIR through grant no.
09/921(0334)/2020-EMR-I. S.C. acknowledges the INSPIRE scholarship from the Department of Science and Technology, Government of India.  CESSI is funded by IISER Kolkata, Ministry of Education, Government of India.

\section*{Data Availability}

We have used 1000-year open solar flux data available at VizieR Online Data Catalog \citep{2021yCat..36490141U}. Data from our simulations will be shared upon reasonable request to the corresponding author.



\bibliographystyle{mnras}
\bibliography{references} 







\bsp	
\label{lastpage}
\end{document}